\documentstyle[aps,pre,multicol,epsf]{revtex}

\def \e {\epsilon^2}
\def \d {\delta}
\def \k {\kappa}
\def \a {\alpha}
\def \L {\Lambda}

\def\BE{\begin{equation}}
\def\BEA{\begin{eqnarray}}
\def\EE{\end{equation}}
\def\EEA{\end{eqnarray}}
\def\B.#1{{\bbox{#1}}}
\begin{document}
\renewcommand{\thesection}{\arabic{section}}
\title{
Anisotropic Spectra of   Acoustic Turbulence}
\author {V.S L'vov$^{*\ddag}$,  Yu.V. 
 L'vov$^\dag$  and A. Pomyalov$^*$}
\address{
$^*$Departments of Chemical Physics, The Weizmann
Institute of Science, Rehovot 76100, Israel,\\ 
$^\dag$ Center for Nonlinear Studies, Los 
Alamos National Laboratory,  Los Alamos, MN 87545,  USA \\
$^\ddag$Institute for Automatization, Russian Ac. of Science, 
Novosibirsk 630090, Russia  } 
\maketitle  
\begin{abstract}
We found universal anizopropic spectra of acoustic turbulence with the
linear dispersion law \bbox{$\omega (k) =ck$} within the framework of
generalized kinetic equation which takes into account the finite time
of three-wave interactions. This anisotropic spectra can assume
both scale-invariant and non scale-invariant form. The implications
for the evolution of the acoustic turbulence with non-isotropic
pumping are discussed.  The main result of the article is that the
spectra of acoustic turbulence tend to become more isotropic.
\end{abstract}
\begin{multicols}{2}
\section{Introduction and General Discussion}
Wave turbulence, which describes the behavior of a spatially
homogeneous field of random dispersive waves, has led to spectacular
success in our understanding of spectral energy transfer processes in
plasmas, oceans and planetary atmospheres
\cite{ZLF}. In the case of small level of nonlinearity $\epsilon$ 
(for example, for the surface waves this is the ratio of the wave
amplitude $h$ to the wavelength $\lambda$, $\epsilon=h/\lambda$),
there is a consistent description of the {\em weak wave turbulence} in
terms of so called {\em kinetic equation} (KE) which describes the
energy transfer due to interactions of three (in some cases four)
waves with the conservation of energy and momenta
\begin{equation}\omega({\B.k})
\pm \omega({\B.k_1}) - \omega({\B.k_2})
=0\,,\quad {\B.k} \pm {\bf
k_1} - {\B.k_2}=0\ .
\label{cons}
\end{equation}
These equations on a classical language are a condition of time-space
resonance. Equations~(\ref{cons}) show that acoustic waves are
special: interacting waves have to have parallel wave vectors:
\BE\label{par}
\B.k \parallel \B.k_1\parallel \B.k_2\,,
\EE
and therefore the interacting acoustic waves are foliated into
noninteracting rays of waves propagating in different directions.
Immediately a few  important questions arise:
\begin{enumerate}
\item What are the mechanisms responsible for an 
 energy redistribution between different rays?
\item How does the energy become shared between neighboring rays?
\item Does the energy tend to diffuse away from the ray with maximum
energy or can it focus onto that ray? In the latter case, one might
argue that shock formation may again become the relevant process
especially if the energy should condense on rays with very different
directions.  And finally:
\item Is the approximation of the KE  adequate for a description of acoustic 
turbulence even in the case of small nonlinearity?
\end{enumerate}
The positive answer on the last question is problematic\cite{ZLF}.
Approximation of KE is based on the randomization of phases of
noninteracting waves leading to the Gaussian statistics and requires
weakness of the interaction $\e\ll 1$ to ensure the closeness the wave
statistics to Gaussianity. This happens in some physical
situations,    but not for acoustic waves.  The physical reason is the
above described foliation: for each particular (noninteracting) ray of
waves one may pass into co-moving (with the sound velocity $c$)
reference system in which all the waves are in the rest and therefore
their interaction time goes to infinity. Therefore for any small level
of nonlinearity the waves have enough time for the finally large
deviation of phases from the random Gaussian distribution. This is
exactly the reason why in one dimensional case one has a creation of
shock waves (which may be described by the Burgers equation) for any
(small) level of nonlinearity. If really acoustic waves tends to
focus, the approximation of KE is problematic even qualitatively. In
this paper we will show that this is not the case and therefore the
approximation of KE may serve at least as a starting point to describe
non-dispersive acoustic turbulence.

The answers on the first three questions for weakly dispersive
acoustic waves was done a long ago by V. L'vov and G. Falkovich\cite{81LF}.
They showed that weakly dispersive waves, say with the dispersion law
\BE\label{weak}
\omega(k)\propto k^{1+\delta}\,,\qquad \delta \ll 1 \,,
\EE
really tends to focus. Namely the  isotropic solution  of the KE
\begin{equation}\label{iso}
n_0(k)=a k^{-9/2}\,,\quad (a\ \mbox{is a dimensional constant})\,,
\end{equation}
found by V. Zakharov and R. Sagdeev in 1970 \cite{ZS} is unstable in the
sense that the anisotropic solutions of the KE found in\cite{81LF}:
\BE\label{LF}
n_{lp}(\B.k)\propto    
\frac{(kL)^{\delta(l+p)/2}}{k^{9/2}}      
\Big[1-\frac{\B.k\cdot\B.n}{k}\Big]^l
\Big[1+\frac{\B.k\cdot\B.n}{k}\Big]^p
\EE
increases with $k$ from the pumping scale $1/L$ toward to the depth of
the inertial interval $kL\gg 1$ faster then the isotropic
one~(\ref{iso}). In (\ref{LF}) $\B.n$ is the unite vector which is
determined by anisotropy of the pumping. Observe  that the ratio
\BE \label{ratio}
\frac {n_{lp}(\B.k)}{n_0(k)}\propto (kL)^{\delta(l+p)/2}
\EE
increases faster for more anisotropic shapes (larger $l$ and $p$) and
for more dispersive waves with bigger $\delta$. In this sense the
non-dispersive waves with $\delta=0$, $\omega(k)\propto k$ are
marginal, $ {n_{lp}(\B.k)}/{n_0(k)} \propto k^0 =$const. and any angle
distribution $n(\B.k)$ is a solution of the KE at $\delta=0$. Clearly
this is a consequence of the foliation described above.  However
physical intuition tells as that there should be some mechanism of
redistributing energy between neighboring rays even for fully
non-dispersive waves. Indeed, the fact that all interacting waves have
parallel wave vectors follows from the time-space resonance
conditions~(\ref{cons}). These conditions are valid with some accuracy
which is determined by the life time of the waves and therefore 
neighboring rays can really interact. In our recent paper\cite{AT}
we generalized the KE for acoustic waves to account the finite width
of the resonances. Generalized Kinetic Equation (GKE) for the
``occupation numbers of waves'' $n(\B.k,t)$ has the form:
\BE\label{GKE}
\frac{\partial n(\B.k,t)}{\partial t}=\mbox{St}_\B.k(\{ n(\B.k',t)\})\,,
\EE
where the collision term $\mbox{St}_\B.k(\{ n(\B.k',t)\})$ is a
functional of the occupation numbers $n(\B.k',t)\}$ with all wave vectors
$\B.k'$ but at the same moment of time $t$ [which for the shortness we
will skip from the arguments: $n(\B.k',t)\Rightarrow n(\B.k')$]. The
collision term for GKE is very similar to that for the KE: it is
proportional to the square of the amplitude of three wave interactions
$V({\B.k},{\B.k}_1,{\B.k}_2)$, bilinear in $n(\B.k')$ and actually
contains one three-dimensional integration:
\end{multicols}
\leftline{-------------------------------------------------------------------------}
\begin{eqnarray} \label{F3}
\mbox{St}_\B.k(\{ n(\B.k')\})
&=&\int\frac{d{\B.k}_1 d{\B.k}_2}{(2\pi)^3}
\Gamma_{\B.k12}
\Big\{ 
\delta({\B.k}-{\B.k}_1-{\B.k}_2) 
\frac{1}{2}\frac{|V({\B.k},{\B.k}_1,{\B.k}_2)|^2 \left[n({\B.
k}_1)n({\B.k}_2)-n({\B.k})[n({\B.k}_1 )+n({\B.k}_2)]\right]} 
   {[\omega({\B.k})-\omega({\B.k}_1)-\omega({\B.k}_2)]^2
+\Gamma^2_{\B.k12}}
\nonumber \\
&+&\delta({\B.k}+{\B.k}_1-{\B.k}_2)   
\frac{|V({\B.k}_2,{\B.k}_1,{\B.k})|^2 
  \left[n({\B.k}_2)[n({\B.k}_1)+n({\B.k})]
 -n({\B.k})n({\B.k}_1)\right]}
   {[\omega({\B.k})+\omega({\B.k}_1
)-\omega_0({\B.k}_2)]^2+\Gamma^2_{\B.k12}}
\Big\} \ .
\label{GGKE}
\end{eqnarray}
\rightline{-------------------------------------------------------------------------}
\begin{multicols}{2}
\noindent
This collision term differ from that of  the KE in the finite width
$\Gamma_{{\B.k} {\bf{k}_1} {\bf{k}_2}} $ of the
resonances~(\ref{cons}):
\BE\label{G3}
\Gamma_{{\B.k} {\B.k_1} {\B.k_2}} 
= \gamma(\B.k)+\gamma(\B.k_1)
+\gamma(\B.k_2)\,,
\end{equation}
where $\gamma(\B.k)$ is the damping of monochromatic wave with given
$\B.k$.  This allows interaction of waves from different, but
neighboring rays.  We will see that for small nonlinearity, $\e \ll 1$
the characteristic angle of interaction is small: $\Delta \theta_k\ll
\pi$. This helps us to significanlty simplify the collision integral.

The paper is written as follows. In Sect.~\ref{s:diff} we simplify the
collision term of the GKE by using the differential in angle
approximation. Observe from Eqs.~(\ref{N2p}, \ref{St2}) and
(\ref{Anizotropic}) below that the collision term St$_\B.k$ in the
differential approximation has just one-dimensional $k$-integration
along the ray with direction of $\B.k$ and operators of
differentiation in two orthogonal directions.  Note, that
Eq.~(\ref{Anizotropic}) written for distributions $n(\B.k)$ with
characteristic angle width much larger then the interaction
angle. This equation has an isotropic solution $n_0(k)\propto
k^{-9/2}$ which coincide with the solution~(\ref{iso}) of the KE.

In Section~\ref{s:linear} we linearize differential form of the GKE
assuming that the deviation $n_1(\B.k)$ of the distribution $n(\B.k)$
from isotropic solution $n_0(k)$ is small and expand 
$$
n_1(\B.k)\equiv n(k,\cos\theta )\,, \qquad 
\cos\theta\equiv \B.k\!\cdot\!\B.n/k
$$
into series of Legendre polynomials $P_\ell(\cos \theta)$:
\begin{equation} \label{n1Leg}
n_1(k,\cos \theta) =\sum_{\ell =1}^\infty 
f_\ell (k)P_\ell(\cos \theta)\ .
\end{equation}
After that our problem foliates into set of decoupled equations for the
Legendre polynomial with a given order $\ell $. In such a way we
reduce the dimensionality of the problem to dimension one. Now the unknown
function $f_\ell (k)$ depends only on one variable $k$ and the 
corresponding collision terms involve only one-dimensional
$k$-integration.  Our observation is that the equations for different
$P_\ell$ involve  only one combination of the parameters, namely $\e
\ell(\ell+1)$.  In Subsection~\ref{ss:sol} we found scale-invariant
solutions of these equations:
\BE\label{scale-inv}
f_\ell (k)\propto \frac{1}{k^{x_\ell}}\,, \quad 
x_\ell = 6+\frac{\ln[\e \ell(\ell+1)B ]}{\ln(k_*L)}\,,
\EE
which are valid for the region of parameters where $5<x_\ell <6 $. In
Eq.~(\ref{scale-inv}) $B$ is some number of order one, $L$ is outer
scale of turbulence and $k_*$ is the wave vector for which
$\gamma(k_*)$ is about the smallest frequency of waves in the system,
$\simeq c/L$.  Note that the isotropic solution~(\ref{iso}) $n_0(k)\propto
k^{-9/2}$ and therefore when $kL$ increases the ratio $ f_\ell
(k)/n_0(k)$ decrease at least as $1/\sqrt{kL}$. It means that in a
cascade of energy transfer from anisotropic region of pumping down to
the depth of the inertial interval energy tends to diffuse between all
the rays and asymptotically, in the limit $k\to \infty$ acoustic
turbulence become fully isotropical. This phenomenon of isotropisation of
non-dispersive acoustic turbulence contrasts with self-focusing of
weakly dispersive acoustic turbulence discovered in\cite{81LF}, and
see Eqs.~(\ref{weak}, \ref{LF}, \ref{ratio}).  Note also that for
non-dispersive waves the rate of isotropization, {\em i.e.} the rate
of decreasing the ration $ f_\ell (k)/n_0(k)$, depends only on the
combination $\e \ell(\ell+1)$ and increases both with $\e$ and $\ell$.

In Section~\ref{s:nonscale} we found non-scale invariant solution of
the linearized equations for $f_\ell(k\L)$ which depends  on some
characteristic length $\L$:
\BE\label{nonscale}
f_\ell(k\L)\sim \sqrt{k\L} \exp \big[-\case{1}{2}
\sqrt{\e \ell(\ell+1)C} (k\L)^2 \big]\ .
\EE
Here $C$ is some constant of order unity. Solution~(\ref{nonscale}) is
valid for $k\L>1$, the value of $\L$ has to be found from a matching
of this solution with a solution for smaller $k$. Again, the rate of
isotropisation depends only on the combination $\e \ell(\ell+1)$ and
increases both with $\e$ and $\ell$.

Note that the choice between two found solution is delicate issue and
depends on the various parameters of the problem: $\e$ and
$\ell(\ell+1)$ separately, on the value of underground dispersion of
the waves, {\em etc.} We do not think that it is reasonable to study this
question in general, without referring to a specific  physical
realization of the acoustic turbulence.

 The main qualitative message of this paper
is that in spite of anisotropic pumping at large scales $L$ acoustic
turbulence became to be more and more isotropic with increasing of its
wave vector.  The rate of isotropisation increases with increasing the
level of nonlinearity $\e$ and depends on the characteristic angle of
the distribution $\Delta \theta \simeq \pi/\ell$ like $1/(\Delta
\theta)^2$. 

\section{Differential Approximation  of \\
the Generalized Kinetic Equation}
\label{s:diff}
Our starting point is the generalized kinetic equation,
GKE~(\ref{GKE}, \ref{GGKE}) which describes the interaction of
neighboring rays due to the finite width of the three-wave resonance
(\ref{G3}), determined by the damping increment of an individual wave
with given $k$, $\gamma(k)$. The value of $\gamma(k)$ was calculated
in\cite{AT}:
\BE\label{E11}
 \gamma(\B.k)=\nu k^2\,,\quad \nu\simeq  \frac{A^2 N } {4 \pi c }\,,
\EE
where $\nu$ is the effective viscosity, $N$ is the  total number of the waves
in the system
\BE\label{N}
N=\int\limits_{1/L}^\infty {n(q) q^2 d q}\,,
\EE
and $A\simeq  \sqrt{c/\rho}$ characterizes the three-wave interaction
amplitude 
\begin{equation}
V({\B.k}, {\B.k}_1, {\B.k}_2)=A\sqrt{k\,k_1\,k_2}\ .
\label{Vkkk}
\end{equation} 
Here we accounted that the interaction is dominant by the interaction of
almost collinear wave vectors and neglected the angular dependence of 
$V({\B.k}, {\B.k}_1, {\B.k}_2)$.

Integrating (\ref{GGKE}) with the help of the delta-functions, one has
\begin{eqnarray} 
\mbox{St}_\B.k(\{n(\B.k')\})
 &=& { A^2 k\over (2\pi)^3}
\Big[ \frac{1}{2} \int d {\B.k}_1 k_1 (k-k_1) 
{ N_- \Gamma_-\over \Omega_-^2 + \Gamma_-^2}   \nonumber\\ 
&& +\int d {\B.k}_1 k_1 (k+k_1) \frac{ N_+\Gamma_+}{\Omega_+^2 + 
\Gamma_+^2}\Big] \ .
\label{KE2} 
\end{eqnarray} 
Here we have used the short-hand notations:
\begin{eqnarray}\label{def2} 
\Gamma_\pm&=&\nu[k^2+k^2_1+|\B.k\pm \B.k_1|^2] , 
\\ \nonumber
N_\pm&=& n(\B.k\pm \B.k_1)[\B.n(k_1)\pm n(\B.k)]  - n(\B.k) 
n(\B.k_1), \\ \nonumber 
\Omega_\pm&=&c(k\pm k_1 -|{\B.k}\pm{\B.k}_1 |)\ .  
\end{eqnarray}
Under the above assumption of dominance of the collinear interaction
one may take ${\bf k}_1 \parallel {\B.k}$ in $\Gamma_\pm$ to get
\begin{equation}\label{gpm}
\Gamma_\pm = 2 \nu (k^2 + k_1^2 \pm k k_1)\ .
\end{equation}
However we cannot use  the same approximation in the Eqs. for
$\Omega_\pm$ and $N_\pm$. To understand this let us have a look at the 
angular integrals in~(\ref{KE2}) at given value of $k_1$.  It is clear
that the main contribution to the integrals comes from the region
where $|\Omega_\pm|$ varies from 0 to a value of order
$\gamma(k)$. Within this region $\Gamma_\pm$ is constant~(\ref{gpm})
with accuracy of order $\gamma(k)/ck\sim \e$. Similar considerations
shows that we have to account a variation of $N_\pm$ within the region
of angular integration in~(\ref{KE2}). 

\begin{figure}
\hskip 1cm
\epsfxsize=  5.truecm
\epsfbox{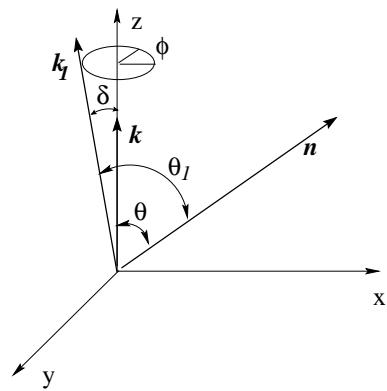}
\narrowtext
\vskip 0.5cm
\caption{The coordinate system used in calculation of the integrals in 
Eq.(7).Vector $\B.n$ denotes the physically unique direction. The axis 
$\B.z$ is along $\B.k$, the axis $\B.x$ is  in the plane ($\B.k$,$\B.n$). 
The angles between vector $\B.n$ and vectors $\B.k$ and $\B.k_1$ are 
denoted by   $\theta$ and $\theta_1$, respectively.  $\d$ is the small 
angle  between  $\B.k$ and $\B.k_1$. The azimuthal angle of
$\B.k_1$  is denoted by $\phi$.}
\label{f-vector}
\end{figure}

To evaluate the integrals in Eq. (\ref{KE2}) one has to establish the
relations between different wave vectors appearing in the equation.
Let $\B.n$ be some physically unique direction and assume for
simplicity an axial symmetry around $\B.n$. The vectors $\B.k$ and
$\B.k_1$ are almost collinear with a small angle $\d$ between them and
arbitrary oriented.  Then we introduce a Cartesian coordinate system
with the axis $\B.z$ along $\B.k$ ( Fig. \ref{f-vector}).
Given the geometry, we now derive a differential approximation for the
anisotropic GKE.  To this end, consider (\ref{def2}) and expand
$\Omega_\pm$ and $N_\pm$ in $\d$ . Clearly,
\begin{equation} 
\Omega_\pm \simeq \d^2 \Omega^{\prime\prime}_\pm\,,
\qquad \Omega^{\prime\prime}_\pm \equiv \pm \frac{ 
c k k_1}{2 (k \pm k_1)} \ ,
\label{Opm1} 
\end{equation} 
whereas $N_\pm$ may be written as:
\begin{equation} N_\pm  = 
N_\pm^{(0)} + N_\pm^{(1)}+ N_{\pm}^{(2)}. \end{equation} 
Here
\begin{equation}
N_\pm^{(0)}
= n_{\B.k\pm \B.q}(n_\B.q\pm n_{\B.k})  - n_\B.k n_\B.q,
\label{def3} \end{equation}
with $\B.q\equiv \B.k k_1/k$.  Terms with $N_\pm^{(1)}
\propto \d \cos\theta$ or  $\d\sin\theta$ and disappear after
integration over $\phi$, while  $N_\pm^{(2)}=\d^2 N_
\pm^{\prime\prime}$.  After (free)
integration over $\phi$, $N_\pm^{\prime\prime}$ reads:
\begin{eqnarray}\label{N2p}
&& N_\pm^{\prime\prime} =
 \frac{1}{4}\Big[ (n_\B.q \pm n_\B.k)
\nabla_\perp^2n_{\B.k\pm\B.k_1}\\
&&+
(n_{\B.k\pm\B.q}-n_{\B. k})\nabla^2_\perp n_{\B.k_1}
\pm
2(\nabla_\perp n_{\B.k_1})
\cdot(\nabla_\perp n_{\B.k\pm\B.k_1})
\Big] \,, \nonumber
\end{eqnarray}
where
\begin{equation}
\nabla_\perp \equiv \sin \theta {\partial \over \partial  \cos \theta_1}
\end{equation}
and $\theta_1$ is the angle between vectors $\B.k_1$ and $\B.n$;
derivatives are taken at $\theta_1=\theta$.

Note, that all the $\d$-dependence of the integrand in Eq. (\ref{KE2})
is hidden in $\Omega_\pm $ and $N_\pm$ and therefore the integrals
may be factorised. Using smallness of $\d$ we write explicitly
$d{\B.k}_1=
\pi q^2 d q d \d^2$ and consider the integrals over $\d^2$:
\begin{equation} I_\pm
\equiv\pi\int d \d^2 \frac{N_{\pm}\Gamma_\pm}{\Omega^2_\pm
+\Gamma_\pm^2} =I_\pm^{(0)}+I_\pm^{(2)} \,,
\label{Ipm}
\end{equation}
where
\begin{eqnarray}
I_{\pm}^{(0)} &=& \pi N_{\pm}^{(0)}\Gamma_\pm
\int \limits_0^\infty
\frac{ d \d^2 }
{\d^4(\Omega^{\prime\prime}_{\pm})^2 + \Gamma_{\pm}^2}
=\frac{\pi^2 }{2}\,\frac{N^{(0)}_\pm }
{ \Omega^{\prime\prime}_\pm}\,,
\label{Ipm1}\\
I_\pm^{(2)}&=&\pi N_\pm^{\prime\prime}\Gamma_\pm
\int \limits_0^b  \frac{\d^2  d \d^2
}{\d^4
(\Omega_\pm^{\prime\prime})^2+\Gamma_\pm^2} =
 \frac{\pi }{2}
\frac{N_\pm^{\prime\prime}\Gamma_\pm {\cal L}_\pm}
{(\Omega_\pm^{\prime\prime})^2}\ .
\label{Ipm2}
\end{eqnarray}
Here the upper limit in the second integral $b$ is determined by the
next terms of expansion of frequency in $\delta$, generally speaking
$b=O(1)$. In Eq.~(\ref{Ipm2}) ${\cal L}_\pm\simeq \ln
(\Omega_\pm^{\prime\prime}/ \Gamma_\pm)$.
Finally, substituting (\ref{Ipm} -- \ref{Ipm2})  into
Eq.~(\ref{KE2}) we get the anisotropic GKE~(\ref{GKE}) with the
collision term in the differential approximation
\begin{equation}
\mbox{St}_\B.k(\{n(\B.k')\})
={\rm St}_{\B. k,0}(\{n(\B.k')\})+ {\rm St}_{\B. k,2}(\{n(\B.k')\})     \ .
\label{collision}
\end{equation}
Here St$_{\B. k,0}(\{n(\B.k')\})$ originates from $N_\pm^{(0)}$ and
coincides with the collision term of KE\cite{ZLF}:
\begin{eqnarray} \label{St0}
\mbox{St}_{\B. k,0}(\{n(\B.k')\})
= \frac{A^2}{8 \pi c} &\Big[&
\frac{1}{2}\int_0^k d q q^2 (k-q)
^2 N_-^{(0)}  \\ \nonumber
&+& \int_0^\infty d q q^2
(k+q)^2 N_+^{(0)} \Big] \ .
\end{eqnarray}
Term St$_{\B. k,2}(\{n(\B.k')\})$ is responsible for the angular
evolution (and disappears for the isotropic distributions of
$n(\B.k)$, as expected):
\begin{eqnarray}\nonumber
\mbox{St}_{\B. k,2}(\{n(\B.k')\})&= &\frac{A^2}{ 8\pi ^2 c}
\Big[
\frac{1}{2} \int_0^k d q q^2 (k-q)^2
\frac{\Gamma_-{\cal L}_-}{\Omega_-^{\prime\prime}}
N^{\prime\prime}_-  \\  \label{St2}
&&+\int_0^\infty d q q^2 (k+q)^2 \frac{\Gamma_+
{\cal L}_+}{\Omega_+^{\prime\prime}}
N_{+}^{\prime\prime} \Big]  \ .
\end{eqnarray}
%%%%%%%%%%%%%%%%%%%%%%%%%%%%%%%%%%%%
\section{Weakly  Anisotropic spectra \\
of acoustic turbulence}
\label{s:linear}
In this Section we will find and analyze a steady state weakly
anisotropic solution to the anisotropic GKE~(\ref{GKE}) with the
collision term (\ref{St0}, \ref{St2}) in  the differential approximation:
\BE\label{Anizotropic}
{\rm St}_{\B. k,0}(\{n(\B.k')\})
+ {\rm St}_{\B. k,2}(\{n(\B.k')\})=0\ .
\EE
\subsection{Linearization of the basic equation}
Due to the weak anisotropy of the problem, we are  seeking a
solution in the form:
\begin{equation} \label{substitution2}
n({\B.k}) = n(k,\cos\theta)=n_0(k) + n_1(k,\cos \theta) \ .
\end{equation}
Here $n_0(k)$ is given by~(\ref{iso}) and is a steady state solution of
the isotropic problem:
\begin{equation}\label{ZS} 
{\rm St}_{\B. k,0}(\{n_0(k')\})=0\ .
\end{equation}
The anisotropic correction is assumed to be small: $n_1(k,\cos
\theta)\ll n_0(k)$.  Then we substitute (\ref{substitution2}) to
(\ref{Anizotropic}) and linearize (\ref{Anizotropic}) by keeping only
terms proportional to $n_1$ and discarding terms with higher orders of
the correction.  To this end, we expand the anisotropic correction
$n_1(k,\cos\theta)$ in a series~(\ref{n1Leg}) in which
$P_\ell(\cos\theta)$ is the Legendre polynomial of the order $m$,
satisfying the equation:
\begin{equation} \label{Peq}
[\nabla^2 _\perp +\ell(\ell+1)]P_\ell[\cos(\theta)]=0 \, .
\end{equation}

Consider first St$_{\B. k,0}(\{n(\B.k')\})$ given by Eq.~(\ref{St0}).
According to~(\ref{ZS}) $n_0 (k)$ is the steady state solution of the
GKE. Therefore the terms proportional to $n_0(k)$ have to disappear.
The result (linear in $n_1$) is given by
\begin{equation} \label{LinearizedST0} 
\mbox{ St}_{\B. k,0}(\{n(\B.k')\}
=\sum_{\ell=1}^{\infty}P_\ell(\cos\theta) \Phi_0(k,f_\ell) \, ,
\end{equation}
where
\end{multicols}
\widetext
\leftline{-------------------------------------------------------------------------}
\begin{eqnarray}\label{f0} && \Phi_0(k, f_\ell)= \frac{A^2}{4 \pi
c} \Bigg[ \frac{1}{2}\int\limits _0^k d q q^2 (k-q)^2 \Big\{
f_\ell(k-q)[n_0(q)-n_0(k)] +n_0(k-q)[f_\ell(q)
-f_\ell(k)] -f_\ell(k) n_0(q) \\
\nonumber &-& n_0(k) f_\ell(q) \Big\} + \int\limits _0^\infty d q q^2
(k+q)^2 \Big\{ f_\ell(k+q)[n_0(q)+n_0(k)] 
+n_0(k+q)[f_\ell(q) +f_\ell(k) ]
-f_\ell(k) n_0(q) - n_0(k) f_\ell(q) \Big\} \Bigg]
\end{eqnarray}
To linearize $ {\rm St}_{\B.k,2}(\{n(\B.k')\})$ we substitute in
Eq.~(\ref{St2}) the distribution $n(\B.k) $ defined by
(\ref{iso}, \ref{n1Leg}, \ref{substitution2}) and using
(\ref{Peq}) to get:
\begin{equation}
\label{LinearizedST2}{\rm St}_{\B. k,2}(\{n(\B.k')\})
=-\sum_{\ell=1}^{\infty}\ell(\ell+1)
P_\ell[\cos(\theta)] \Phi_2(k, f_\ell)\,, 
\end{equation}
where
\begin{equation}
\Phi_2(k, f_\ell)= \frac{A^2}{ 8\pi ^2 c}
\Big(
\frac{1}{2} \int_0^k d q q^2 (k-q)^2
\frac{\Gamma_-{\cal L}_-}{\Omega_-^{\prime\prime}}
N^{\prime\prime(\ell)}_-
+\int_0^\infty d q q^2 (k+q)^2 \frac{\Gamma_+
{\cal L}_+}{\Omega_+^{\prime\prime}}
N_{+}^{\prime\prime(\ell)} \Big) \ ,
\label{St2p}
\end{equation}
and
$$
N^{\prime\prime(\ell)}_\pm = \frac{1}{4}\Big\{ \big[n_0 (q) \pm n_0 (k)
\big]
 f_\ell(k\pm q)+\big[n_0(k\pm q)-n_0(k) \big] f_\ell(q) \Big\}\ .
$$
%%%%%%%%%%%%WWWWWWWWW
\rightline{---------------------------------------------------------------
----------}
\begin{multicols}{2}
The steady state weakly anisotropic GKE therefore reads:
\begin{equation}
\label{GKEp}
\sum_{\ell=1}^\infty \Big[ \Phi_0(k, f_\ell) -\ell(\ell+1)\Phi_2(k, f_\ell)
\Big]= 0 \ .
\end{equation}
The particular solution of this equation which may satisfy any
boundary conditions may be found only if each term in the sum
vanishes, i.e.
\begin{equation}
\label{GKEm}
\Phi_0(k, f_\ell) =\ell(\ell+1)\Phi_2(k, f_\ell)
 \ .
\end{equation}
This is the basic equation for our study.

\subsection{Evaluation of  the collision integral}
\label{ss:coll1}
In order to solve Eq.~(\ref{GKEm}) in the leading order on the class
of scale invariant functions
\begin{equation}\label{sol1}
f_\ell(q)=\frac{\phi_\ell}{q^x}\ ,\quad \phi_\ell\ 
\mbox{is a prefactor}\,,
\end{equation}
one has to evaluate the integrals in $\Phi_0$ and $\Phi_2$ and to find
a leading contribution in the various regions of the for exponents
$x$.  These is done in Appendix~\ref{a:A}, here we only will present
and discuss the results.

As we show in Appendix~\ref{a:P0}, the leading contributions to
$\Phi_0$~(\ref{f0}) may be written as
\begin{equation}\label{IR+UV}
\Phi_0 \approx  \frac{A^2   n_0(k)\phi_\ell(2x-9)}{2
\pi c  | (x-4)(x-5)|} \left\{
\begin{array}{cc}
 L^{x-5}\,,  & \quad    x> 5 \,, \\
                     &                          \\
k^{5-x}\,,  & \quad    4< x<5 \,, \\
                     &                          \\
 k \, k_*^{x-4} \,, & \quad    x< 4 \ .  \\
\end{array}
\right.
\end{equation}
For $x>3$ two integrals in~(\ref{f0}) diverge in the IR regime (for
$q\to 0$ and for $q\to k$). However the leading divergent terms are
canceled and the region of IR divergence becomes $x>4$. Moreover,
first subleading terms also canceled and real region of IR divergence
in~(\ref{f0}) is $x>5$. In this regime one has to cutoff the integral
at outer scale $L$, see first line in
Eq.~(\ref{IR+UV}). For $4<x<5$ the sum of integrals in~(\ref{f0})
converge both in IR and ultraviolet ({\em i.e}. for $q\gg k$)
regimes. The corresponding evaluation is given by the second line in
Eq.~(\ref{IR+UV}). Finally, for $x<4$ the value of $\Phi_0$ is
dominated by the ultraviolet (UV) divergent contribution to the second
integral in~(\ref{f0}) and have to be regularised by some UV cutoff
$k_*$.  Corresponding evaluation is given by the last line
in~(\ref{IR+UV}).  The origin of $k_*$ is different for different
physical system and will not be discussed here. Observe from equation~(\ref{IR+UV})  that (i)  $\Phi_0$ has definite sign, (ii) $\Phi_0=0$ at x=9/2, 
consistent with \cite{ZLF}, (iii) prefactor in $\Phi_0$ diverges for
$x\to 5$ and $x\to 4$.

In Appendix~\ref{a:P2} we evaluated the leading contributions to
integrals in Eq.~(\ref{St2p}) for $\Phi_2$ for different values of
$x$. The answers may be summarized as follows:
\begin{eqnarray}\label{UV+IRest}
&& \Phi_2(k, f_\ell)\approx
\frac{A^2\nu{\cal L} n_0(k)\phi_\ell }{ 2 \pi ^2 c^2}\\ \nonumber
&\times & \left\{ \begin{array}{cc}
k^3 L^{x-3}&\mbox{( for}\quad 6<x ) \,,\\
&\\
k^3 L^{x-3}+k_*^{6-x}&\mbox{for}\quad \case{11}{2}<x<6\,,\\
&\\
-(kL)^{5/2}k^{6-x}
+k_*^{6-x}&\mbox{for}\quad  x <\case{11}{2}\ .\\
\end{array} \right.
\end{eqnarray}
In the IR regime we have two different contribution to $\Phi_2$, the
first is divergent for any $x$, the second is divergent for
$x>3$. These contributions coincide at $x=11/2$.  One may recognize
the contributions from the IR regimes by the presence of $L$ in the
corresponding terms.  In the integral (\ref{St2p}) for $\Phi_2$ we
have UV-divergence for $x<6$, in this region the upper cutoff again is
$k_*$ and corresponding terms involve this parameter. Note that in
contrast to evaluations~(\ref{IR+UV}) we do not have here a window of
locality where the collision integrals converge.

\subsection{Solution of the Homogeneous Equation}

\label{ss:sol}
\begin{figure}
\epsfxsize=  8.5truecm
\epsfbox{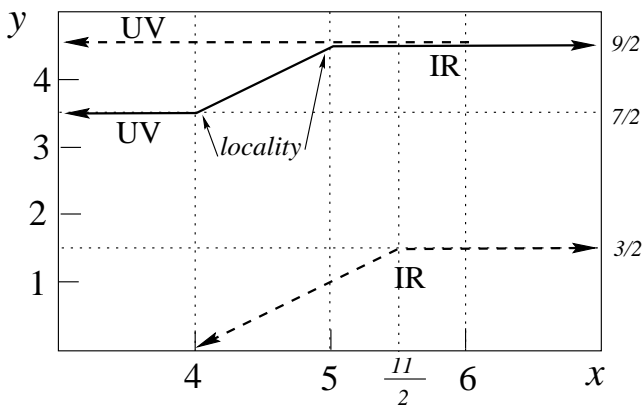}
%\narrowtext
\vskip 0.5cm
\caption{
Scaling exponents $y_0$ (solid line) and $y_2$ (dashed lines) for
different values of x. Arrows show the directions of divergence for
the corresponding integrals. ``IR'' and ``UV'' stay for the infrared and
ultraviolet limits, ``locality'' denotes the region of convergence of
$\Phi_0$.}
\label{xy-fig}
\end{figure}

Our goal  now is to find the solution of the homogeneous Eq.~(\ref{GKEm}).
We  seek the solution in the scale invariant sector.
Denote the scaling exponents of $\Phi_0(k, f_\ell)$ and  $\Phi_2(k, f_\ell)$
as $y_0$ and $y_2$:
\begin{equation}\label{exp-y02}
\Phi_0(k, f_\ell)\propto \phi_\ell k^{-y_0}\,,\quad
\Phi_2(k, f_\ell)\propto\phi_\ell k^{-y_2} \ .
\end{equation}
The $x$ dependences of these exponents are determined by
Eqs.~(\ref{IR+UV}, \ref{UV+IRest}) and shown  in Fig. \ref{xy-fig}.

$\Phi_0$ converge for $4<x<5$ and diverge for other values of $x$ (in
IR for $5<x$ and in UV for $x<4$).  $\Phi_2$ never converge with
different rate of divergence in IR and UV limits (two branches, see
Fig. \ref{xy-fig}).  The ratio between IR and UV-divergent terms in
$\Phi_2$ depends on the wave vector $k$, therefore, generally speaking
we have to account both IR  and UV terms.
\narrowtext

Fortunately, the exponents $y_0=y_2=\case{9}{2}$ coincide in the
window $5<x<6$, meaning that at least the $k$-dependence of
$\Phi_0(k,f_\ell)$ and $\Phi_2(k, f_\ell)$ is the same.  Next promising
observation is the coincidence of sings: both function are
positive. It means that we have a chance to satisfy Eq.~(\ref{GKEm})
by a proper choice of the exponent $x$. Indeed, according to
(\ref{IR+UV},~\ref{UV+IRest}), Eq.~(\ref{GKEm}) for these values of
$x$ becomes:
\begin{equation}\label{hom1}
L^{x-5}=B'\, (x-5)\,\ell(\ell+1)\frac{\nu k_*^{6-x}}{c}\,,
\end{equation}
where $B'$ is some positive, $\ell$-independent, dimensionless factor
[we believe of $O(1)$] which accumulated all unknown factors in our
estimates.

Let us estimate effective viscosity $\nu$. Substituting isotropic
solution~(\ref{iso}) in~(\ref{N}) one gets $N\simeq aL^{3/2}$. Now
from~(\ref{E11}) one has 
\BE\label{nu1}
\nu\simeq aL^{3/2}/\rho\ .
\EE
 Next one has to relate the dimensional factor $a$ with the
 dimensionless amplitude of waves $\epsilon$. To do this we evaluate
 the total energy of the acoustic waves
\BE\label{energy}
E_{\rm ac}=\int d^3k \, \omega(k)n_0(k)\simeq 4\pi \,c \,a \!\! 
\int\limits_{1/L}^\infty
\frac{dk}{k^{3/2}}\simeq c\,a\sqrt{L}\ .
\EE
We  define $\e$ as the the ratio of the total energy in the
acoustic-wave system $E_{\rm ac}$ to the total kinetic energy of the
media $E_{\rm kin}$ which is about $\rho\, c^2$. Parameter $\epsilon$
may be treated as the ratio of the amplitude of velocity in acoustic
waves to the mean square velocity in the media caused by its kinetic
energy in the thermodynamic equilibrium.

Now we have $\e\simeq a\sqrt{L}/\rho\,c$ which together
with~(\ref{nu1}) gives
\BE\label{nu2}
\nu\simeq \e c\,L\ .
\EE
Using this evaluation in Eq.~(\ref{hom1}) one has:
\begin{equation}\label{hom2}
L^{x-6}=B^{\prime\prime}\e  \,\ell(\ell+1)k_*^{6-x} \,,
\end{equation}
where $B^{\prime\prime}$ absorb one more unknown factor from the
evaluation of $\nu$.  Therefore the solution is achieved for
 $x=x_{0,\ell}$, where
\begin{equation}\label{hom3}
x_{0,\ell}= 6+\frac{ \ln[\e \ell(\ell+1)B^{\prime\prime}]
}{\ln(k_*\, L)}\ .
\end{equation}
These exponents give the $k$-dependence of the solution (\ref{GKEm})
 in the depth of the inertial range. However,
the functions $\phi_\ell$ remain unknown. To find them one have to
match the inertial range solution to the pumping at the IR boundary
($k=1/L$).

\subsection{Effect of  Inhomogeneous Terms}
Sometimes the inhomogeneous terms in the KE causes additional solutions
which may play important role in the evolution of spectra in
$k$. Consider first the origin of the inhomogeneous terms. Linearizing
Eq.~(\ref{LinearizedST0}) we have concluded that $n_0^2$ contribution
to St$_{\B.k,0}=0$ is zero. This is true only if the IR limit of the
integral is indeed zero, which is not the case. In the energy
containing interval $0<k<L^{-1}$ we have a non-universal behavior of
$n({\B.k})$ which has to be accounted. A simple way to do this is
first to evaluate the contribution of this region as
\begin{equation}\label{inhomog1}
\mbox{St}_{\B.k,0,\rm inhom}(\{n(\B.k'\})\simeq
\frac{A^2\,n_0({k})  }{2\pi c \sqrt{L}} n \Big(\frac{\B.k}{k\,L}\Big) \ .
\end{equation}
Note that in this region $n({\B.k})$ is not isotropic and thus
   St$_{\B.k,0,\rm inhom}(\{n(\B.k'\})$ must depend on the direction
   of $\B.k$.  Expanding this dependence into the spherical harmonics
   (in our case of the axial symmetry of the problem into the Legendre
   polynomials) we have an inhomogeneous contribution to the
   Eq.~(\ref{GKEm}):
\begin{equation} \label{inhomog2}
\Phi_{\rm inhom}(k,n_{0,\ell})\approx \frac{A^2 n_0(k) n_{0,\ell}}{2 \pi c
\sqrt{L}} \propto k^z\,,
\end{equation}
where non-universal numbers $n_{0,\ell}$ may be related to $\phi_\ell$
via ``boundary conditions'' at $k\approx 1/L$:
\begin{equation} \label{inhomog3}
n_{0,\ell}\approx L^{x-9/2}\phi_\ell\ .
\end{equation}
An important observation is that the scaling exponent $z$ of
$\Phi_{\rm inhom}(k,n_{0,\ell})$ in~(\ref{inhomog2}) is independent of
$x$ (because this term is independent of $f_\ell(k)$
altogether). Moreover, $z=\frac{9}{2}$, which is exactly the same as
exponents $y_0$ and $y_2$ of the homogeneous part of the GKE $\Phi_0$
and $\Phi_2$.  It means that for any $k$ one has to account the
contribution of $\Phi_{\rm inhom}(k,n_{0,\ell})$ in the balance
equation~(\ref{GKEm}).  Repeating the same calculations as in deriving
Eq.~(\ref{hom1}) with the help of Eqs. (\ref{GKEm}, \ref{IR+UV})
and~(\ref{UV+IRest}) and accounting now evaluations~(\ref{inhomog2},
\ref{inhomog3}) we are approaching again Eq.~(\ref{hom1}) with
additional contribution to the factor $B^{\prime\prime}$; their sum
may be denoted as $B$.  Physically it means that resonant (i.e. with
the same scaling exponent) inhomogeneous term has shifted the
exponents $x_\ell$ relative to their ``homogeneous
values''$x_{0,\ell}$~(\ref{hom3}). By replacing the unknown constants
$B^{\prime\prime}\Longrightarrow B$ we obtained from~(\ref{hom3})
final  equation~(\ref{scale-inv}) for the scaling exponents  of $x_\ell$.

\section{Non-scale invariant Solutions}
\label{s:nonscale}
In previous Sect.  we found all scale-invariant solutions and observe
that they have finite region of applicability for which their scaling
exponents $5<x_\ell<6$. Outside of this region we have to look for 
non-scale invariant solutions which contain explicitly a
characteristic length-scale $\L$ which will be chosen later.   To do
that we have to choose $f_\ell(k\L)$ in some reasonable form, for
example
\begin{eqnarray} \label{sol2} f_\ell(\k
)&=&a_{0,\ell}\k ^x\left[ 1+ \frac{\a_{2,\ell}}
{ \k^2} + \frac{\a_{4,\ell}}{
\k^4 }\dots \right] \\ \nonumber 
&\times &\exp \big[ - b_\ell \k^y\big]
\,,\quad \k\equiv k\, \L \ge 1\,, 
\end{eqnarray} 
where $x$, $y$, $a_{i,\ell}$ and $b_\ell$ are some unknown numbers.
Obviously there is no UV   divergences in $\Phi_0$ and $\Phi_2$ for
such choice of $f_\ell$. Furthermore, since in {\it leading order}
$\lim\limits_{q\to 0} f_\ell=a_0$ there is no IR divergence in
$\Phi_0$ and $\Phi_2$ associated with $f_\ell$, so that the IR
behavior of the $\Phi_0$ and $\Phi_2$ is dominated by 
$n_0(q_{_{\rm IR}}),\ \ \
q_{_{\rm IR}}\sim 1/L$.

Then in expression (\ref{f0}) we keep only the terms proportional to
$n_0(q)$ and $n_0(k-q)$:
\end{multicols}
\widetext
\leftline{----------------------------------------------------------------------------}
\begin{eqnarray}\label{f0nons1}
\Phi_0(k, f_\ell)&\simeq &\frac{A^2}{4 \pi c} \Big[
\frac{1}{2}\int\limits _{q_{_{\rm IR}}}^k d q q^2 (k-q)^2 
\Big\{ f_\ell(k-q)n_0(q)
+n_0(k-q)[f_\ell(q)-f_\ell(k)] -f_\ell(k) n_0(q) \Big\} 
\\ \nonumber
&& + \int\limits _ {q_{_{\rm IR}}}
d q q^2 (k+q)^2 \Big\{ f_\ell(k+q)n_0(q)
-f_\ell(k) n_0(q)
\Big\} \Big]\ .
\end{eqnarray}
Now let us change variables in the first integral $q\to k-q$. Then we
can evaluate integrals at $q=q_{_{\rm IR}}$ only:
\begin{eqnarray}\nonumber &&\Phi_0(k, f_\ell) 
\simeq \frac{A^2}{4 \pi c}
\Big[
\frac{1}{2}\int\limits _ {q_{_{\rm IR}}}d q q^2 (k-q)^2 
\Big\{ f_\ell(k-q)n_0(q)
+n_0(k-q)[f_\ell(q)-f_\ell(k)] -f_\ell(k) n_0(q) 
+f_\ell(q)n_0(k-q) \\ \label{f0nons2} 
&& +n_0(q )[f_\ell(k-q)-f_\ell(k)]-f_\ell(k-q) n_0(k-q) \Big\} +
\int\limits _ {q_{_{\rm IR}}}d q q^2 (k+q)^2 
\Big\{ f_\ell(k+q)n_0(q) -f_\ell(k) n_0(q)
\Big\} \Big]\simeq\nonumber\\  
&&\simeq \frac{A^2}{4 \pi c} \Bigg[
\frac{1}{2}\int\limits _ {q_{_{\rm IR}}}d q q^2 k^2 2 
n_0(q )[f_\ell(k+q)-2 f_\ell(k) +
f_\ell(k-q)] \Big]  \simeq \frac{A^2}{8 \pi c} \frac{k^2
\partial^2 f_\ell(k)}{\partial k^2} 
\int\limits _{q_{_{\rm IR}}} d q q^4 n_0(q ) \ .
\end{eqnarray}
\rightline{-----------------------------------------------------------------------------}
\begin{multicols}{2}
\noindent
The second derivative of $f_\ell$ is the result of the  usual  double
cancellation of the IR divergence in the isotropical part of KE.

Now we have to evaluate $\Phi_2$ at $q\ll k$ for the choice of $f_\ell$
given by (\ref{sol2}). Let us first evaluate $N^{\prime\prime}_-$ for 
$q\ll k$:
\begin{equation}
N^{\prime\prime(\ell)}_- \! \simeq \case{1}{4}\big\{ n_0 (q) f_\ell(k-q)
 +\big[n_0(k- q)-n_0(k) \big] f_\ell(q) \big\},
\end{equation}
or
\begin{equation}
N^{\prime\prime(\ell)}_- \simeq
\case{1}{4} n_0 (q)  f_\ell(k), \ \  q\ll k
\end{equation}
Similarly,
\begin{equation} 
N^{\prime\prime(\ell)}_+ \simeq
\case{1}{4} n_0 (q) f_\ell(k),\ \ \  q\ll k.
\end{equation} Now use (\ref{est0}) to
estimate $\Phi_2$: 
\begin{eqnarray} \label{St2p2} \Phi_2(k, f_\ell)&
\simeq & \frac{A^2}{ 16 \pi ^2 c} \int\limits_{q_{_{\rm IR}}} 
d q q^2 k^2 \frac{4 \nu
k^2\,{\tilde{\cal L}}}{c q} n_0 (q) f_\ell(k) \\ \nonumber && \simeq
\frac{A^2}{ 16 \pi ^2 c} q_{_{\rm IR}}^2 k^4 4 \e L 
n_0(q_{_{\rm IR}}) f_\ell(k),\ \ \ 
 q_{_{\rm IR}}\simeq 1/L\,,
\end{eqnarray} 
where we used estimation (\ref{nu2}) for effective viscosity 
$\nu$.  Actually we have not calculated accurately the $q\to k$
contribution in $\Phi_2$. Note however, that this contribution would
be of lower (in $q$) order, because $ \lim_{q\to k}\simeq -[c k ^2/ 2
(k - q)]$.

Now Eq.~(\ref{GKEm}) acquires the form (introducing dimensionless
constant $C$, generally speaking of the order of unity)
\begin{equation}
\label{GKEmnonscale}
q_{_{\rm IR}}^5 k^2 n_0(q_{_{\rm IR}} )\frac{d^2 f_\ell(k)}
{d k^2} = C\e
L \ell(\ell+1) q_{_{\rm IR}}^2 k^4 n_0(q_{_{\rm IR}}) f_\ell(k) \ .  
\end{equation} 
Note that $n_0(q_{_{\rm IR}})$ cancels from both sides of this
equation. Further notice that both terms are of positive sign.
Canceling $q_{_{\rm IR}}$ and $k$ from both sides and substituting
$\kappa=k/\Lambda$ with $q_{_{\rm IR}}\simeq 1/L$ we get
\begin{equation}
\label{schema} \frac{d^2 f_\ell(\k)}
{d \k^2} \simeq C \ell(\ell+1) \e \k^2 f_\ell(\k) \,, \quad \k\equiv
k\,L\ ,  \end{equation}
where we chose $\L\simeq L$.
 This equation may be solved in the Bessel
functions (of order $\case{1}{4}$ of imaginary argument). For our
goals, however, it would be enough to present only its asymptotic form
for $\k \gg 1$: 
\begin{equation}\label{sol3} f_\ell(\k)
=\sqrt{\k}\Big[1+ \frac{a_{2,m}}{\k^2}+\dots \Big]
\exp\Big(-\frac{\epsilon}{2}\sqrt{C \ell(\ell+1)}\k^2\Big )\,, 
\end{equation}
where coefficients $a_{2,\ell}$, $a_{4,\ell}$, {\em etc}. may be found
iteratively: $a_{2,\ell}=3/ \sqrt{C \ell(\ell+1) \e}$, $\dots$ but these terms
really are beyond the accuracy of Eq.~(\ref{schema})
itself. Solution~(\ref{sol3}) shows that anisotropic corrections decay
exponentially, the rate of decay increases with $\e$ and $\ell(\ell+1)$.

\section{Summary} Let us summarize the logic of this paper:
\begin{itemize}
\item We start from Generalized Kinetic Equation (GKE)  (\ref{GGKE}),
\item{The conditions of time-space resonances~(\ref{cons})
 dictates {\it almost} collinear propagation, so we expand
 (\ref{GGKE}) in the transverse direction, and arrive to the
 differential approximation for the collision term in GKE 
 (\ref{collision}) with (\ref{St0}) and (\ref{St2})};
\item We assume {\it weak} anisotropy with anisotropic correction in
the factorized form (\ref{substitution2});
\item After  substitution  (\ref{substitution2}) in Eqs.~ (\ref{collision}, 
\ref{St0}, \ref{St2}), we  {\it linearize} resulting 
three-dimensional equations (which are differential in angles and
integral along the rays);
\item Expansion of these equations in series in Legendre polynomials
 $P_\ell$ foliates the 3-dimensional problem in decoupled sets on
 1-dimensional ones, for each $\ell$ separately;
\item We have found the power-law solutions of this
equations~(\ref{scale-inv}) and the exponential
solutions~(\ref{nonscale}) which are governed by the same parameter
$\e \ell (\ell +1)$. Thses solutions describe the phenomenon of
isotropization of non-dispersive acoustic turbulence in the cascade
process of energy transfer from anisotropic region of pumping down to
the smaller and smaller scales ($k\to\infty$). In the limit
$k\to\infty$) the statistics of acoustic turbulence approaches
isotropical flux equilibrium.
\end{itemize}
The main result of the article is that the spectra of acoustic
turbulence tend to become more isotropic.

\acknowledgements
This work has been supported in part by the Israel Science Foundation
administered by the Israel Academy of Sciences, and the Naftali and
Anna Backenroth-Bronicki Fund for Research in Chaos and Complexity.

\appendix

\section{       Analysis of converges \\
of integrals in the collision terms}
\label{a:A}
\subsection{$\Phi_0(k,f_\ell)$ behavior}
\label{a:P0}
\subsubsection{IR-regime} 

In Eq.~(\ref{f0}) one has three different regions dangerous for  IR
divergence. These are the regions $q\to 0$ and $q\to k$ in the first
integral and $q\to 0$ in the second one. The first integral
$\case{1}{2}\int_0^k\dots$ may be split into two integrals $\case{1}{2}
\int_0^{k/2}\dots$ and 
$\case{1}{2}\int_{k/2}^k\dots$  In the second one we may change dummy
variable $q\to q'=(k-q)$ and then to re-denote $q'\to q$. After that
one sees that the integral $\case{1}{2}\int_{k/2}^k\dots$ equal to
$\case{1}{2} \int_0^{k/2}\dots$ and therefore the first integral
in~(\ref{f0}), $\case{1}{2}\int_0^k\dots$, may be rewritten as
$\int_0^{k/2}\dots$.  This integral in the region $q\ll k$ together with
the second integral in~(\ref{f0}) may be written as
\begin{eqnarray}\label{f0IR}
&& \Phi_0(k,f_\ell) \approx \frac{A^2}{4 \pi c}\int_0^{k/2}d q q
^2(k-q)^2 f_\ell(q) \\ \nonumber &\times&[n_0(k-q) -
n_0(q)]+(k+q)^2f_\ell(q) [n_0(k+q)-n(k)] \ .
\end{eqnarray}
Substituting $f_\ell(q)$ from~(\ref{sol1}) one observes here a usual double
cancellation of the leading and the first subleading terms in the IR
regime.  Therefore integral~(\ref{f0IR}) may be evaluated (up to a
factor) as follows:
\begin{equation}
\Phi_0(k,f_\ell) \approx \frac{A^2\,n_0(k)}{4 \pi c}
\int\limits _0^{k/2}d q q^4 f_\ell(q)\ .
\label{Phi0eval}
\end{equation}
Now one sees that the integral (\ref{Phi0eval}) converges for $x<5$
and diverges for $x>5$.  In the later case we take the outer scale $L$
as an IR cutoff. Then it is easy to see that $\Phi_0$ may be evaluated
in different regions of $x$ as:
\begin{equation}\label{IR}
\Phi_0 \approx \frac{A^2 n_0(k)\phi_\ell}{2 \pi c } \left\{
\begin{array}{cc}
 k^{5-x}/ (5-x) & \quad \mbox{(for }x<5) \,, \\ & \\ \ln (kL) & \quad
                     \mbox{(at }x= 5)\,, \\ & \\ L^{x-5}/ (x-5) &
                     \quad \mbox{(for } 5< x) \ .  \\
\end{array}
\right.
\end{equation}
Note, that in this limit and for $x$ close to 5, $\Phi_0>0$.
\subsubsection{UV-regime}
 Next we analyze the region $q\gg k$, {\em  i.e.}  the UV-regime in the
 second integral in Eq.~(\ref{f0}).  Here the most dangerous terms
 reduce to
\begin{equation}\label{f0UV}
\Phi_0(k,f_\ell) \approx \frac{A^2\, n_0(k) }
{4 \pi c}\int \limits _k^{k_*}d q q
^4  \big[ f_\ell(k+q)-f_\ell(q)\big]\ .
\end{equation}
There is a cancellation of the leading contribution in the region
$q\gg k$.  Therefore this integral converges at $x>4$. In a compact
form the UV-evaluation of $\Phi_0$ may be written as:
\begin{equation}\label{UV}
\Phi_0 \approx - \frac{A^2 k n_0(k)\phi_\ell}{2 \pi c | x-4 |} \left\{
\begin{array}{cc}
 k^{4-x}\,,  & \quad    x > 4 \,, \\
                     &                          \\
 k_*^{4-x} \,, & \quad    x <  4 \ .  \\
\end{array}
\right.
\end{equation}
Here $k_*$ is an UV cutoff of the integral which may have different
nature for different physical situations and will not be further
clarified here.  Note, that in the UV-regime for $x$ close to 4,
$\Phi_0< 0$.  It is known also, that $\Phi_0= 0$ for $x=9/2$, the
scaling index of the isotropic solution of the KE. Now we can join
Eqs.~(\ref{IR}) and (\ref{UV}) and write evaluation of the leading
contribution to $\Phi_0= 0$ in all regions of $x$ in the
form~(\ref{IR+UV}).

\subsection{$\Phi_2(k,f_\ell)$ behavior}\label{a:P2}
\subsubsection{IR-regime}  
Unlike $\Phi_0(k,f_\ell)$ , there is no double cancellation of the IR
contribution ({\em i.e}. in the limits $q \to 0$ and $q\to k$) in the
sum of the integrals~(\ref{St2p}) for $\Phi_2(k,f_\ell)$. The reason
is that $\Omega''_\pm$ is not invariant under $q\to k-q$
transformation.  Now, in the IR region $q\ll k$ one has the following
simplifications: 
\begin{eqnarray}\label{est0}
\Gamma_\pm &=& 2 \nu (k^2 + k_1^2 
\pm k k_1) \simeq 2 \nu k^2\, , \\
\nonumber \Omega_\pm &\simeq &\d^2 
\Omega^{\prime\prime}_\pm\,, \qquad
\Omega^{\prime\prime}_\pm \equiv \pm 
\frac{ c k k_1}{2 (k \pm k_1)}
\simeq \pm c q /2 \ , \\ \nonumber {\cal L}_\pm&=& \ln
\Big(\frac{\Omega_\pm^{\prime\prime}}
{ \Gamma_\pm} \Big) \simeq \ln
\Big(\frac {c q } {4 \nu k^2}\Big) 
\equiv \tilde {\cal L}\ .
\end{eqnarray}
Substituting these equations into (\ref{St2p}) one gets in the IR region:
\begin{equation}\label{est1}
\Phi_2(k, f_\ell)\simeq \frac{\nu k^4 A^2
\tilde {\cal L}}{ 2 \pi ^2 c^2 }
\int\limits _0^{k} d q q \left[\frac{1}{2} N^{\prime\prime(\ell)}_- -
N^{\prime\prime(\ell)}_+ \right] \ .
\end{equation}
 The most dangerous terms in the combinations
 $N^{\prime\prime(\ell)}_\pm $ are:
\begin{equation}
N^{\prime\prime(\ell)}_\pm \simeq
\frac{1}{4}\big\{ n_0 (q)
 f_\ell(k\pm q)+\big[n_0(k\pm q)-n_0(k) \big] f_\ell(q) \big\}\ .
\end{equation}
Together with   Eq.~(\ref{ZS}) these yields:
\begin{equation}
N^{\prime\prime(\ell)}_\pm \simeq
\frac{1}{4}\Big\{ n_0 (q)
 f_\ell(k)\mp \frac{9 \, q}{2 \, k} n_0(k)  f_\ell(q) \Big\}\,.
\end{equation}

Then in the IR limit  $\Phi_2$ reads
\begin{eqnarray}\label{est2}
\Phi_2(k, f_\ell)& \simeq & \frac{\nu k^4 
A^2\tilde {\cal L}}{\phi ^2 c^2
} \int \limits _0^{k} d q\, q\\ \nonumber &\times & \left[-
n_0(q)f_\ell(k) + \frac{q }{\,k} n_0(k) f_\ell(q) \right] \,,
\end{eqnarray}
where we did not care about the numerical factor, just carrying the
signs.  Since $n_0(k)\propto k^{-9/2}$ the $n_0(q)f_\ell(k)$ term
always diverges; the corresponding contribution to the integral 
behaves as $(kL)^{5/2} k^2 n_0(k)f_\ell(k)$. The second term will
diverge if $f_\ell(q) = \phi_\ell/q^x$ with $x<3$. Symbolically, its
contribution to the integral are:
\begin{equation}\label{est3}
\int _0 \dots \sim \frac {n_0(k)\, \phi_\ell}{k\, |x-3|}
\left\{
\begin{array}{cc}
   k^{3-x} & \quad \mbox{( for}\quad x< 3)\,, \\ & \\ L^{x-3 } & \quad
\mbox{( for}\quad x>3) \ .  \\
\end{array}
\right.
\end{equation}
Observe that for $x>11/2$ the contribution of the second term in
(\ref{est2}) dominates the contribution of the first one. Finally, the
IR contributions to $\Phi_2 (k, f_\ell)$ are summarized as:
\BEA \label{IRest}
\Phi_2   (k, f_\ell) &\approx &\frac{\nu k^3\, A^2 {\cal L}n_0(k)\phi_\ell}
{2 \pi^2 c^2}\\ \nonumber
&&\times
\left\{
\begin{array}{cc}
  - (kL)^{5/2}k^{3-x}   &\mbox{( for}\quad x<11/2)\,,  \\
                     &                            \\
 L^{x-3 } &\mbox{( for}\quad x> 11/2) \ .\\
\end{array}
\right.
\EEA
\subsubsection{UV-regime} 
Next we have to consider the UV regime of the integral in~(\ref{St2p})
{\em i.e.} the region of integration with $q\to k_\ell \gg k $. Here
we have the simplifications:
\begin{eqnarray}\label{est4}
\Gamma_+ &=& 2 \nu (k^2 + k_1^2 \pm k k_1) \simeq 2 \nu q^2\,,
\\ \nonumber
 \Omega_+ &\simeq&  \d^2 \Omega^{\prime\prime}_\pm\,,
\qquad \Omega^{\prime\prime}_\pm \equiv \pm
\frac{ 
c k k_1}{2 (k \pm k_1)} \simeq \frac {ck} {2} \ , \\ \nonumber 
{\cal L}_+ &= & \ln  \Big( \frac{\Omega_\pm^{\prime\prime}}
{\Gamma_\pm} \Big )  \simeq \ln \Big( \frac {c k  }
{4 \nu q^2}\Big )  \equiv \hat {\cal L} \ .
\end{eqnarray}
Now $\Phi_2$ becomes 
\BEA
\Phi_2(k, f_\ell)&=& \frac{A^2}{ 8\pi ^2 c}
\int \limits _k^{k_*} d q q^2 (k+q)^2 \frac{\Gamma_+ 
{\cal L}_+}{\Omega_+^{\prime\prime}} 
N_{+}^{\prime\prime(\ell)}\\  \nonumber
&\simeq&
\frac{A^2\nu \hat {\cal L}
}{ 2 \pi ^2 c^2 k}
\int \limits _k^{k_*} d q q^6  N_{+}^{\prime\prime(\ell)} \ .
\EEA
Out of $N^{\prime\prime(\ell)}_+$ the most dangerous for UV term is
\begin{equation}
N^{\prime\prime(\ell) }_+\simeq  \frac{1}{4}\big[n_0 (k)( 
 f_\ell(k+ q) - f_\ell(q)) \big] \sim -\frac{x\, k}{4\, q}n_0(k)f_\ell(q)\,,
\end{equation} 
where we used that $ f_\ell(q)\propto q^{-x}$ and $q\gg k$. Now in the
UV regime the $\Phi_2(k, f_\ell)$ term may be evaluated (up to a
numerical factor) as
\begin{equation}
\Phi_2(k, f_\ell)\simeq  
- \frac{A^2\nu\hat {\cal L} n_0(k)}{ 2 \pi ^2 c^2 }
\int \limits _k^{k_*} d q q^5  f_\ell(q)\ .
\end{equation}
This integral converges for $x>6$ and may be written as follows:
\begin{equation}\label{UVest}
\Phi_2(k, f_\ell)\sim
- \frac{A^2\nu\hat {\cal L} n_0(k)\phi_\ell }{ 2 \pi ^2 c^2 |x-6|}
\left\{ \begin{array}{cc}
k_*^{6-x}&\mbox{for}\quad x<6\,,\\
&\\
k^{6-x}&\mbox{for}\quad x>6\ .\\
\end{array} \right. 
\end{equation}
Combining Eqs.~(\ref{UVest}) and ~(\ref{IRest}), one has after some
minor manipulations (neglecting factors of order one, difference
between ${\cal L}$ and $\hat{\cal L}$, {\em etc}.)
Eq.~(\ref{UV+IRest}).

\end{multicols}

\begin{thebibliography}{99}
\bibitem{ZLF} V.E. Zakharov, V.S.L'vov and  G.Falkovich,
   {\em Kolmogorov spectra of turbulence I.}, v.1, (Springer-Verlag,
1992).

\bibitem{81LF} V.S. L'vov and G.E. Falkovich.  Zh. Eksp. Teor.
      Fiz. {\B.80},592 (1981), [Sov. Phys. JETP, 53(2), Feb. 1981]. 

\bibitem{ZS}  V.E. Zakharov and R.Z. Sagdeev, Sov. Phys. Dokl. {\B.15},

\bibitem{AT} V.S.Lvov, Y.V.Lvov, A.C.Newell and  V.E.Zakharov, PRE,
{\B.56}, 390 (1997)
\end{thebibliography}
\end{document}